\documentclass[12pt, a4paper]{iopart}

\usepackage[square, comma, sort&compress, numbers]{natbib}
\usepackage{graphicx}
\usepackage{circuitikz}
\usepackage{booktabs}
\usepackage{epsfig}
\usepackage{subfigure}
\usepackage{fixltx2e}
\usepackage{amssymb}
\usepackage{wasysym}
\usepackage[margin=10pt,font=footnotesize,labelfont=bf,labelsep=period]{caption}

\def\newblock{}

\newcommand{\caion}{$^{40}$Ca$^{+}$ }
\newcommand{\deptheff}{$\kappa_{\mathrm{d}}$ }
\newcommand{\vadj}{$V_{\mathrm{a}}$}
\newcommand{\vnom}{$V_{\mathrm{nom}}$}

\begin{document}

\title[2D Arrays of RF Ion Traps with Addressable Interactions]{Two-Dimensional Arrays of RF Ion Traps with Addressable Interactions}

\author{Muir Kumph$^1$, Michael Brownnutt$^1$, and Rainer Blatt$^{1,2}$}

\address{$^1$ Institut f\"ur Experimentalphysik Technikerstr 25/4, 6020 Innsbruck, Austria}
\address{$^2$ Institut f\"ur Quantenoptik und Quanteninformation der \"Osterreichischen Akademie der Wissenschaften, Technikerstrasse 21a, A-6020 Innsbruck, Austria}
\ead{muir.kumph@uibk.ac.at}
\begin{abstract}
We describe the advantages of 2-dimensional, addressable arrays of spherical Paul traps. They would provide for the ability to address and tailor the interaction strengths of trapped objects in 2D and could establish a valuable new tool for quantum information processing. Simulations of trapping ions are compared to first tests using printed circuit board trap arrays loaded with dust particles. Pair-wise
interactions in the array are addressed by means of an adjustable radio-frequency (RF)
electrode shared between trapping sites. By attenuating this RF electrode potential, neighboring pairs of trapped objects have their interaction strength increase and are moved closer to one another. In
the limit of the adjustable electrode being held at RF ground, the two formerly spherical
traps are merged into one linear Paul trap.

\end{abstract}

\maketitle

\section{Introduction: going from ion strings to 2D arrays}

The use of ion traps for trapping, cooling, and performing quantum experiments with ions is well documented~\cite{Blatt:2008, Haffner:2008, Leibfried:2003}.  Typically each ion encodes a single qubit and the experiments are carried out in linear Paul traps.  Such traps have the ability to store a string of ions and, together with coherent radiation sources, allow the creation of entangled states and the implementation of quantum operations.  In a single linear trap, entanglement of up to 14 qubits has been created~\cite{Monz:2010}.  However, for useful, large scale quantum simulation and computing, thousands or even millions of qubits would be needed, which cannot be done in a traditional linear Paul trap.  One possible path to scale up a linear ion trap is by the use of segmented electrodes \cite{Kielpinski:2002}.  The ions would then be shuttled around a segmented microscopic ion trap between various zones.  Increasingly complex electrode structures, capable of shuttling ions between traps, have been produced with the aim of developing a fully functional and scalable quantum computer or quantum simulator~\cite{Amini:2010, Chiaverini:2005, Huber:2008, Pearson:2006, Rowe:2002, Blakestad:2009, Walther:2011}. 

An alternative proposal for scaling up quantum system with ion traps was made by Cirac and Zoller~\cite{cirac:2000} and uses a closely-spaced 2-dimensional array of ion traps.  A highly detuned standing wave, which can create a state dependent force~\cite{cirac:2000}, or a M{\o}lmer-S{\o}rensen-type controlled-phase-gate~\cite{Sorensen:2000} could be used to create an entangled state between ions in neighboring traps.  In both cases the Coulomb force between the ions would be the interaction allowing for the entanglement to be generated~\cite{Harlander:2010, Brown:2011}.  Entangled states could be created between neighboring ion traps, possibly allowing for large scale measurement based quantum computing~\cite{Raussendorf:2003} and simulations of quantum systems~\cite{Buluta:2009}.

\begin{figure}
	\center{\includegraphics[scale=0.4]{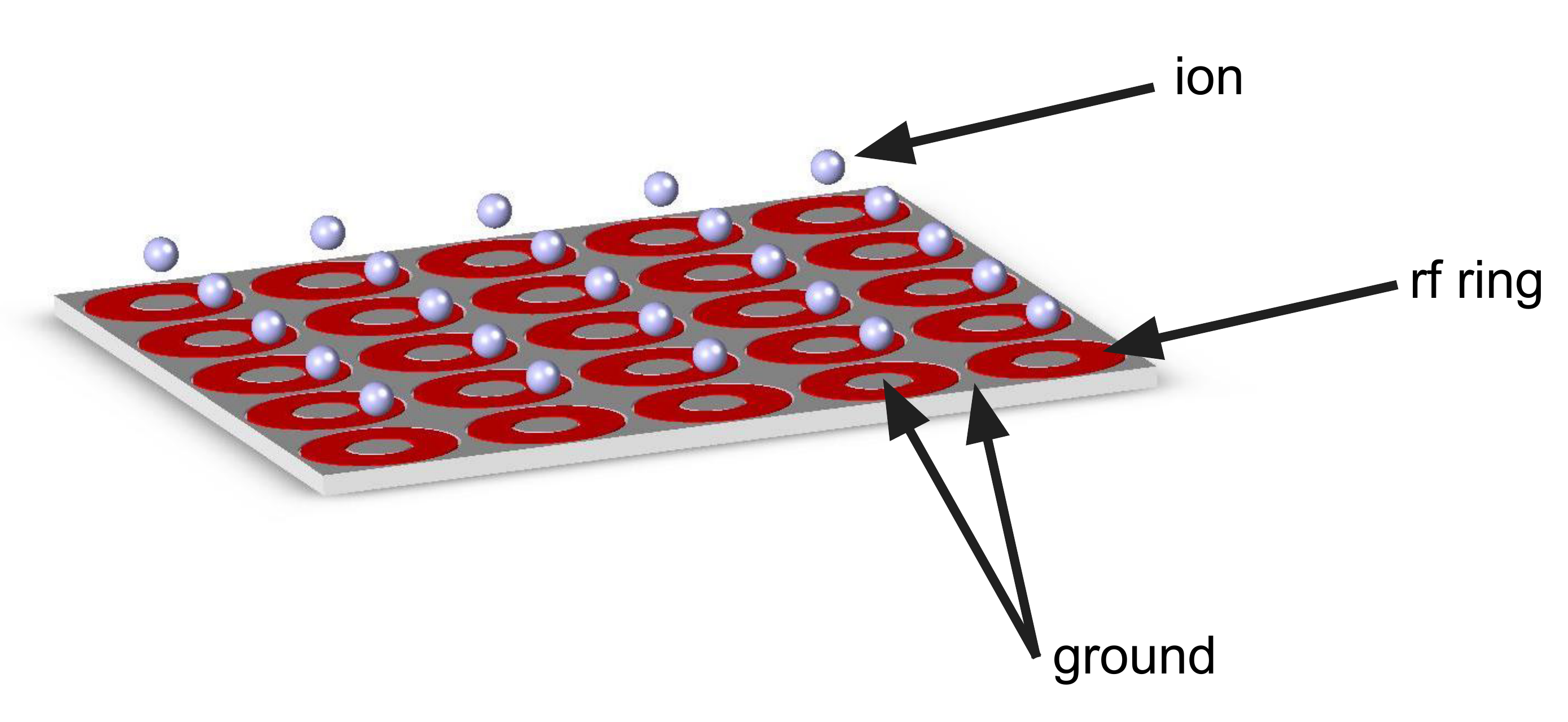}}
	\caption{Above is a representation of how a 2-dimensional 5$\times$5 array of planar point ion traps might appear.  Each ring has a radio frequency voltage applied to it.  Inside each ring is a circular ground electrode.  Outside the ring a ground potential can be applied or a distant ground such as the vacuum chamber can be used.  The ions are trapped above the circular ground electrode.}
	\label{fig:normal2Darray}
\end{figure}

Efforts to use Penning traps~\cite{Hellwig:2010, Biercuk:2009, Castrejon-Pita:2007} or optical dipole forces~\cite{Schneider:2010} to create a periodic 2D array of such micro ion traps are also underway.  Two-dimensional arrays of RF Paul traps~\cite{Clark:2009} have been tested and proposals have been made to improve the trapping dynamics of such arrays~\cite{Schmied:2009} and increase the variety of physical systems which could be simulated~\cite{Welzel:2011}.  Trap arrays can be fabricated with a planar geometry~\cite{Chiaverini:2005, Pearson:2006}, where each trap in the array is referred to as a \emph{point} trap~\cite{Kim:2010}.  Planar structures are desirable in that they allow the use of photolithographic techniques for fabrication.  This permits miniaturization of the traps and  scaling-up of the array through replication.  An important feature of these arrays of point traps is that each trap uses the time varying (typically radio frequency) electric field to confine the ion in all 3 dimensions (see figure~\ref{fig:normal2Darray}).  Planar ion traps which have cylindrical symmetry~\cite{Wesenberg:2008} containing a center trapping site, typically at ground, a radio frequency (RF) ring electrode and a far field ground have been studied as elements which could be arranged in an array~\cite{Schmied:2009}. 

There are however several shortcomings with the above proposals for arrays of Paul traps, which have made them difficult to realize.  One reason is the trade-off between the separation of ions in the neighboring trap sites and the separation between the ions and the electrodes.  For appreciable ion-ion coupling, the proposal of Cirac and Zoller~\cite{cirac:2000} calls for bringing ions within tens of micrometers (or less) of each other.  For most ion-trap array geometries this requires that the ions are also tens of micrometers from the electrodes.  Unfortunately, if the ions are this close to the electrode materials, then motional heating of the ions becomes very high~\cite{Leibrandt:2006, Turchette:2000} compared to the ion-ion coupling.   The motional heating then causes decoherence to dominate and destroys any chance of performing quantum experiments.   Furthermore, if the array of ion traps is designed so that the ions are close to each other, but far from the trap electrodes, the trapping potential falls off dramatically~\cite{Schmied:2009}, making such a trap impractical to load and operate. This is in contrast to ions in segmented linear traps, where their segmented DC electrodes can allow multiple potential wells to be separated by distances smaller than the trap feature sizes~\cite{Harlander:2010, Brown:2011}, since the ions are shuttled along the line of RF null.  Ions are ideally kept at the RF null of a point Paul trap so that they avoid being driven by the trap's RF electric fields.  The point-like nature of the RF null in the 2D arrays of point Paul traps means that ions cannot be significantly displaced from the RF null positions.  Keeping the ions far away from electrode materials, so as to avoid motional decoherence~\cite{Leibrandt:2006, Turchette:2000},  but close enough for an entangling gate~\cite{cirac:2000}, while maintaining an operable trap depth~\cite{Schmied:2009} does not appear particularly feasible in the 2D arrays described above.  

2D arrays of planar-electrode Paul traps with dynamically-reconfigurable RF voltages have been suggested~\cite{Chiaverini:2008, Lybarger:2010} in which ions are shuttled around to control their interactions.  These proposals allow for ions to be shuttled in 2 dimensions, and kept on the RF null.  However, there are a number of significant technical challenges to be overcome.  Others have changed the RF parameters in dust traps~\cite{Karin:2010} or ion traps~\cite{Herskind:2009, Kim:2011} to control the position of the trapping site. This paper proposes a fixed grid topology in which the ion-ion interactions can be \emph{addressed} using variable RF voltages and thereby increased, while maintaining ions in a deep trapping potential, far from the electrodes. The feasibility of the proposed scheme is explored in a dust trap, and solutions to some of the technical challenges for a similar ion-trap system are outlined.

In section \ref{sec:consid} the physics of 2-dimensional arrays of ion traps are described. The results from the simulation of an addressable 2$\times$2 array using variable-voltage-amplitude RF electrodes are given in section \ref{sec:sims}. Section \ref{sec:dust} then presents the experimental results with charged dust particles in the same geometry and similar variable RF voltage. For trapping \caion ions, the design of a miniaturized 4$\times$4 trap is described (section \ref{sec:folsom}) along with the electronics to drive this trap (section \ref{sec:trapdrive}).

\section{Considerations for 2D trap arrays}
\label{sec:consid}
To better explain why using a variable-amplitude RF voltage electrode helps in the goals of 2D ion trap arrays, it helps to explore the physics of 2D trap arrays and consider what are the key features which make them desirable for quantum physics experiments.

\subsection{The competing physical effects of 2D arrays}
Attempts at making ion trap arrays where the Coulomb interaction between ions is strong enough, while at the same time keeping the ions cold and maintaining a deep trap are often stymied by the physics of 2D trap arrays.  There are three characteristics which need to be optimized in a 2D array of ion traps intended for quantum experiments:
\begin{itemize}
\item a strong Coulomb interaction
\item a deep enough trapping potential
\item a low heating rate
\end{itemize}
The first characteristic can be achieved by miniaturizing the array.  The subsequent heating rate caused by the close proximity of the electrodes can be reduced by operating the trap in a cryogenic environment~\cite{Labaziewicz:2008}.  Although the results of heating rate suppression in a cryostat are impressive, the predicted~\cite{Turchette:2000} and measured~\cite{Seidelin:2006} heating rates in traps as small as tens of micrometers could be too high for an experiment with an array of ion traps.  For this reason, it is very desirable to find a way to trap the ions close to each other but still far from the electrodes, even if a cryostat is at hand.  A fourth item to consider is that of trap stability.

\subsection{Trap stability}
In order for an RF Paul trap to stably hold ions, it must be operated within the so-called trap stability regime~\cite{Leibfried:2003}.  To operate near the center of the stability regime, the ratio of the secular frequency $\omega$ to the trap drive frequency $\Omega$ should be approximately 1/7th (see section II. in~\cite{Leibfried:2003}).  Smaller ratios are permissible with attention to DC fields in the trap, but larger values can easily become unstable~\cite{Alheit:1996}.  This requirement can be written as~\cite{Paul:1990},
\begin{equation}
\label{eqn:stability}
\frac{\omega}{\Omega}=\frac{q \kappa V}{K m \Omega^2 d^2}\sim \frac{1}{7}
\end{equation}
where $q$ is the charge of the ion, $m$ is the mass of the ion, $\omega$ is the secular frequency of the ion in the trap, $\kappa$ is the trap efficiency, $d$ is the ion-electrode spacing, $K$ is a constant of order 1 depending upon the principal axis considered, $V$ is the amplitude of the sinusoidal time-varying trap-drive voltage, and $\Omega$ is the frequency (typically RF) of the trap drive voltage.   The trap efficiency $\kappa$ depends upon the geometry of the electrodes.  When the trap electrodes are of a perfect 3-dimensional hyperbolic shape, $\kappa$ is equal to unity~\cite{Paul:1990}, but for planar electrode structures, studied herein, it can be much lower, typically $\sim$0.2.

\subsection{Coulomb interaction between the ions}
The time for an entangling gate between two ions located in separate ion traps given by Cirac and Zoller ~\cite{cirac:2000} requires that the state dependent force be equal to that which displaces the ion the size of the trap ground-state.  With this relatively weak state-dependent force, the gate is operated in a so-called \emph{stiff-mode}~\cite{Porras:2004} regime, where the interaction strength has a dipolar decay law.  The time for a controlled-phase-gate operation is then,
\begin{equation}
T_{gate}=4\pi^2\epsilon_{0} ma^{3} \omega /q^{2}
\label{eqn:gatetime}
\end{equation}
where $\epsilon_{0}$ is the vacuum permittivity and $a$ is the inter-ion spacing.  By reducing the distance between the ions, the time for an entangling gate drops dramatically.  Also by reducing the trap frequency, the gate time can be reduced.  The charge and mass of the ion are considered fixed parameters in our analysis.  If gate operations utilizing stronger state dependent forces could be employed~\cite{Madsen:2006} then the gate times could be significantly reduced.  However, for the analysis presented here, it is assumed that only gate operations employing relatively weak state-dependent forces are used.  

\subsection{Trap depth}
The trap should have a depth capable of trapping ions from a hot oven (a typical atomic source) and holding them long enough to perform useful experiments.  Typical experiments are performed in ultra high vacuum, though collisions do occur with the background gas.  In addition, the heating rate of the ions, which rises as the trap is miniaturized, reduces the trapping time when cooling lasers are turned off.  Experiments with micro ion traps (ion electrode separation $\sim$100~$\mu$m)~\cite{Leibrandt:2009} suggest that at typical pressures (10$^{-9}$ to 10$^{-10}$~mbar), single ion traps with a minimum depth of tens of meV can perform experiments.  

The pseudo-potential, $\Phi$, is given by~\cite{Schmied:2009},
\begin{equation}
\label{eqn:pot}
\Phi=\frac{q^2 \|E\|^2}{4 m \Omega^2}
\end{equation}
where $E$ is the electric field. The trap depth $D$ is the maximum pseudo-potential $\Phi$ experienced by an ion along the lowest-energy escape path (or terminating at an electrode).  Equation~\ref{eqn:pot} is useful for calculating the trap depth when the electric field has been calculated, as done below, with help from a finite-element electrostatic solver.   However, for comparison it is useful to substitute the trap frequency $\omega$, in equation \ref{eqn:stability}, into Hooke's law ($D=m\omega^2d^2/2$) to calculate the trap depth of an optimal hyperbolic electrode trap (axial direction, $K=\sqrt{2}$).  For any quadrupole trap, the trap depth is then~\cite{Wesenberg:2008},
\begin{equation}
\label{eqn:depth}
D=\frac{\kappa_{\mathrm{d}} q^2 V^2}{4 m \Omega^2 d^2}
\end{equation}
where trap depth efficiency \deptheff is defined as the ratio of the trap depth of a given trap to the trapping depth of a perfect hyperbolic-electrode Paul trap.  For a planar-electrode point Paul trap, as proposed in this paper, this efficiency is at most 2 percent~\cite{Wesenberg:2008}, though this can be improved by a factor of 3-4 by simply placing a ground plane a small distance above the array~\cite{Clark:2009}, as has  been done in the analysis presented here.

\subsection{Heating rate of ions as a trap is scaled}
Heating of the ions can limit the performance of entangling gate operations between ions, especially in microtraps~\cite{Seidelin:2006}.  It can cause ion loss when the cooling lasers are turned off.  It is therefore important to consider the scaling of the heating compared to the gate operation time as the trap is miniaturized.

When the constraints of trap stability, a specific trap depth, and fixed trap shape are taken into account, it becomes necessary to increase the trap frequencies $\omega$ and $\Omega$ linearly as the geometry is miniaturized ($\omega \propto 1/d$).   This can be seen by examining the constraint equations (\ref{eqn:stability}) and (\ref{eqn:depth}).   From equation~(\ref{eqn:depth}), a fixed $D$ and \deptheff, requires that $V\propto\Omega d$.  Substituting this into equation~(\ref{eqn:stability}) it is required that $\omega \propto 1/d$.  For an array constructed with relatively high \deptheff planar-electrode, point Paul traps, the inter-ion distance $a$ is related to the ion-electrode spacing $d$ ($a\ge2d$)~\cite{Schmied:2009}.   When $\omega$ is eliminated in equation~(\ref{eqn:gatetime}), the time for an entangling gate then scales as the square of the inter-ion distance $a^2$ instead of $a^3$.  Because the inter-ion distance $a$ and ion-electrode distance $d$ are proportional, it also follows that, as the array is miniaturized, the heating rate goes up.  With array miniaturization, the gate time decreases as $a^2$, but the heating rates have been reported to increase typically with $d^{-4}$~\cite{Turchette:2000}.  The trap array cannot simply be miniaturized to improve performance, since the heating increases faster than the gate time decreases.  Instead, the trap array should be made small enough, but no smaller, so that an entangling gate can be performed, subject to other sources of decoherence.  For laser sources which perform gate operations on \caion ions, high fidelity gates take typically on the order of 50~$\mu$s~\cite{Benhelm:2008b}, though gate operations, which take as long as several hundred $\mu$s, have been performed~\cite{Haffner:2005}.  2-D arrays of ions that are miniaturized enough so that an entangling gate operation takes on the order of $\sim$50-500~$\mu$s would allow for quantum experiments, but further miniaturization would only decrease the fidelity of the gate, because of increased heating.

\subsection{Gate times between traps with a trapping potential of 1 eV}
\label{sec:gatetimes}
Because the voltage applied to the electrodes is limited in magnitude by the breakdown voltage of surface traps (several hundred volts), the relations of equations (\ref{eqn:stability}), (\ref{eqn:gatetime}) and (\ref{eqn:depth}) can be used to calculate the minimum time for an entangling gate (utilizing relatively weak state-dependent forces~\cite{cirac:2000}) as a function of ion-ion spacing in an array of microtraps (for a given \deptheff and $D$).  By including the constraints of a practical planar ion trap ($D\sim$1~eV and \deptheff$\sim5\%$), the gate times shown in the third column of table~\ref{tab:gatetimes} are significantly longer than those initially suggested by Cirac and Zoller~\cite{cirac:2000}, and greatly exceed times for high fidelity gates with \caion ions using known techniques.  In the last column, the trap potential has been relaxed to $\sim$10~meV, resulting in gate times which could conceivably be used for entangling gate operations.  In this relaxed trap, the trap frequency $\omega$ is much less than $1/7^{th}$ of the drive frequency $\Omega$, but trap stability can still be maintained with careful attention to applied DC biases.  Unfortunately, a 2D array of ion traps with a 10~meV trap depth will be of only limited use because of a short experimentation time limited by ion loss.  Frequent reloading of the trap is made all the more difficult by the large number of trapping sites.  Because of these limitations, there are a number of significant issues in creating an experiment with a conventional 2D array of planar ion traps.

\Table{\label{tab:gatetimes}Example gate times for an entangling gate between ions in a 2-dimensional array for various regimes of trap operation.  In the third column are the gate times if the trap depth is $\sim$1~eV.  In the last column, the trap has been relaxed by lowering the trap drive voltage so that the trap depth $D$ is $\sim$10~meV, while the reduced trap frequency $\omega\prime$ is $\omega/10$.}
    \br
    a ($\mu$m) & $\omega$(MHz) &T$_{\mathrm{gate}}$(ms)& reduced T$_{\mathrm{gate}}$(ms) \\
	 & &$D\sim$1 eV &$\omega\prime=\omega/10$, $D\sim$10~meV\\
    \midrule
    1500 & 0.5 & 2200 & 220 \\
    375 & 2 & 140 & 14 \\
    100 & 7.5 & 9.6 & 0.96 \\
    50 & 15 & 2.1 & 0.21 \\
    25 & 30 & 1.3 & 0.13 \\
    \bottomrule
    \end{tabular}
\end{indented}
\end{table}
\subsection{Varying the RF drive amplitude to increase the interaction between ions}
\label{sec:varyRF}
As can be seen in table~\ref{tab:gatetimes}, if the trap potential were lowered to 10 meV (for example by reducing the RF drive voltage amplitude), then useful entangling gates between neighboring ions in a 2D array could be realized in traps with a spacing of 25 $\mu$m. Unfortunately, if the trapping potential of the whole array were this low, then ions would be easily lost during the experiment, making such an experiment very difficult.  However, if only the \emph{pair} of neighboring ions of interest could be \emph{addressed} with a lower RF drive amplitude, so that \emph{their} interaction was increased, it might be possible to do experiments with an array of 2D ion traps.

\section{Geometry and simulations of addressable arrays of ion traps}
\label{sec:sims}
To individually adjust (here called \emph{address}) the interaction between local pairs of point ion traps in an array, an independently adjustable RF electrode between the point Paul traps is inserted.  By attenuating the RF voltage on this \emph{addressing} electrode, the electric field in between the two point Paul traps is reduced.  In this way, a neighboring pair of ions in the 2D array can have their interaction increased, leaving the rest of the 2D array of ions isolated and with a deep trapping potential.   The attenuation of the RF drive voltage on the addressing electrode proposed here similarly allows the ion to be transported~\cite{Karin:2010, Kim:2011}.  It also permits a larger overall trap-depth when compared to lowering the RF drive of an entire array.   Additionally, it allows for the addressing of just pairs of nearest neighbors, by bringing just the pair closer to each other and lowering their shared trap frequency.  The dynamically reconfigurable arrays proposed by Chiaverini and Lybarger~\cite{Chiaverini:2008, Lybarger:2010} would switch a large array of RF electrode pixels ($\sim$25 adjustable electrodes/ion) to allow for a complete reconfiguration of a planar trap providing for the ions to be shuttled between processing zones. What is proposed here is more basic ($\sim$2 adjustable electrodes/ion) and allows for the addressing and adjustment of the interaction between nearest neighbors, where the overall topology of the trap array is not changed.  The proposal here also involves a novel adjustable RF drive, which keeps the phase of the RF voltage the same while the amplitude is adjusted.  The electronics to do this are discussed below in section \ref{sec:trapdrive}.

For analysis, a 2D array of ion traps with an addressing RF electrode between the trapping sites was designed.  This was then simulated using finite-element analysis to compute the electrical field.  The pseudo-potential of the trap was then computed with equation~(\ref{eqn:pot}) to determine the trap depth $D$.  Because these traps are harmonic in the region close to the RF null, the trap frequency, and therefore the trap stability was estimated by fitting a parabolic function to the pseudo-potential $\Phi$. 

\subsection{Addressable Arrays}
\begin{figure}
\center{\includegraphics[scale=0.75]{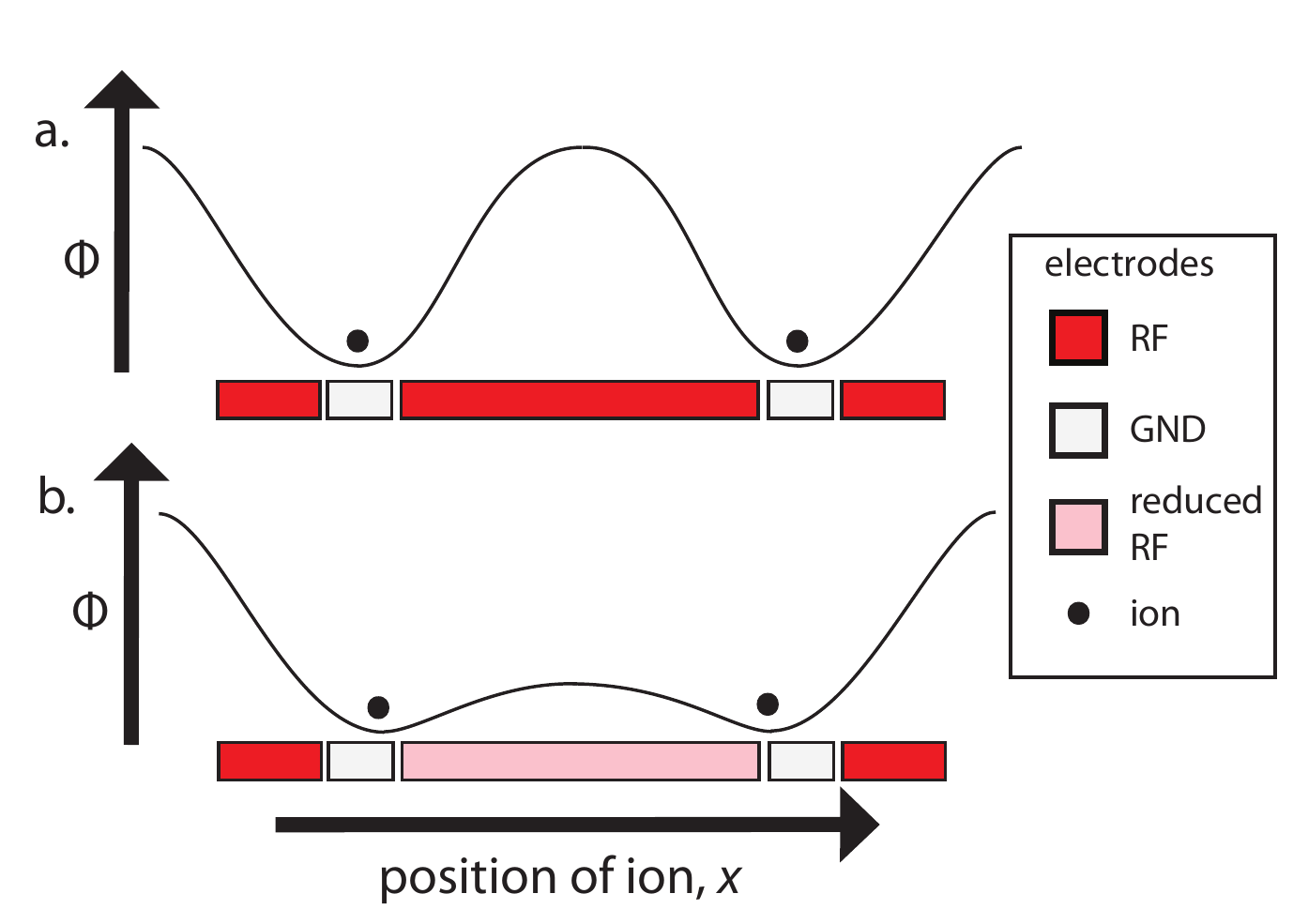}}
\caption{(a) schematic of the trapping potential $\Phi$ between two sites in an ion trap array, and (b) where the potential barrier between the two trapping sites is lowered to increase the relative Coulomb interaction between them.}
\label{fig:doublewell}
\end{figure}

The double-well pseudo-potential along a line connecting two neighboring trapping sites in an addressable 2D array of point Paul traps, is shown in figure \ref{fig:doublewell}.  In figure \ref{fig:doublewell}a is the pseudo-potential of an array when all RF electrodes have the same amplitude, and in figure \ref{fig:doublewell}b is the pseudo-potential when an addressable RF electrode in between the two trapping sites has its RF amplitude reduced.  The interaction between two neighboring ions is increased by modifying the double-well potential such that the central barrier, which separates the neighboring ions, is attenuated.  When the addressable RF electrode has its voltage amplitude lowered, the electric field in between the two trapping sites is reduced.  The pseudo-potential in between the sites and also the trap frequency $\omega$ is then reduced, decreasing the time for an entangling gate operation~(equation~(\ref{eqn:gatetime})).  Due to the subsequent asymmetry of the potential, the ion spacing $a$ is also reduced during this process, further decreasing the time required for an entangling gate operation.  In segmented linear traps, which have an extended RF null,  a similar configuration has been created with the application of DC control voltages~\cite{Harlander:2010, Brown:2011} in 1D.  However, because an RF quadrupole, which creates the pseudo-potential, is confining in at least 2 dimensions, to vary the interaction in more than just one dimension, it is required to modify the RF trap drive field.

\subsection{From static 2D arrays to addressable arrays}
\label{sec:geo}
\begin{figure}
	\center{\includegraphics[scale=0.4]{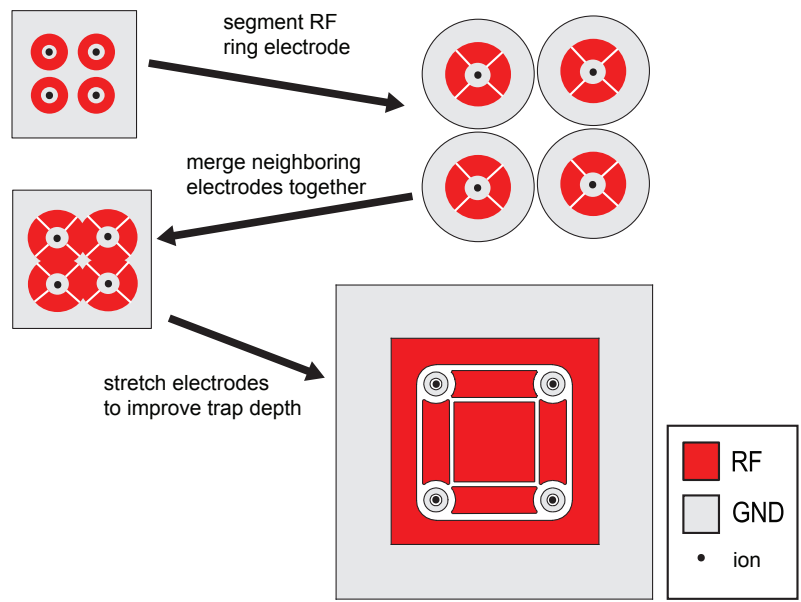}}
	\caption{A representation of the conceptual process in going from a static 2-dimensional array of ion traps to an array where the neighboring ion-ion interaction can be addressed via the shared addressing RF electrode.  Ions are trapped over each circular ground electrode when all RF electrodes are driven with equal voltage.}
	\label{fig:addressable-electrodes}
\end{figure}

Shown in figure~\ref{fig:addressable-electrodes} is the conceptual progression going from a 2$\times$2 static array of 2D traps to an addressable array.  Starting from a normal array of point Paul traps with a planar electrode geometry, each RF ring is segmented.  These rings are then merged together, so that neighboring traps share a segmented addressable RF electrode.  The geometry of both point Paul traps~\cite{Wesenberg:2008} and arrays of such traps~\cite{Schmied:2009} has been extensively studied to optimize various parameters.  Following on this research, the addressing RF electrode's length is increased, so that the trap-depth efficiency \deptheff is improved.  Increasing the addressable electrode's length requires the addition of a fixed RF electrode inside the 2$\times$2 array.  

\begin{figure}
	\centering
	\includegraphics[scale=0.3]{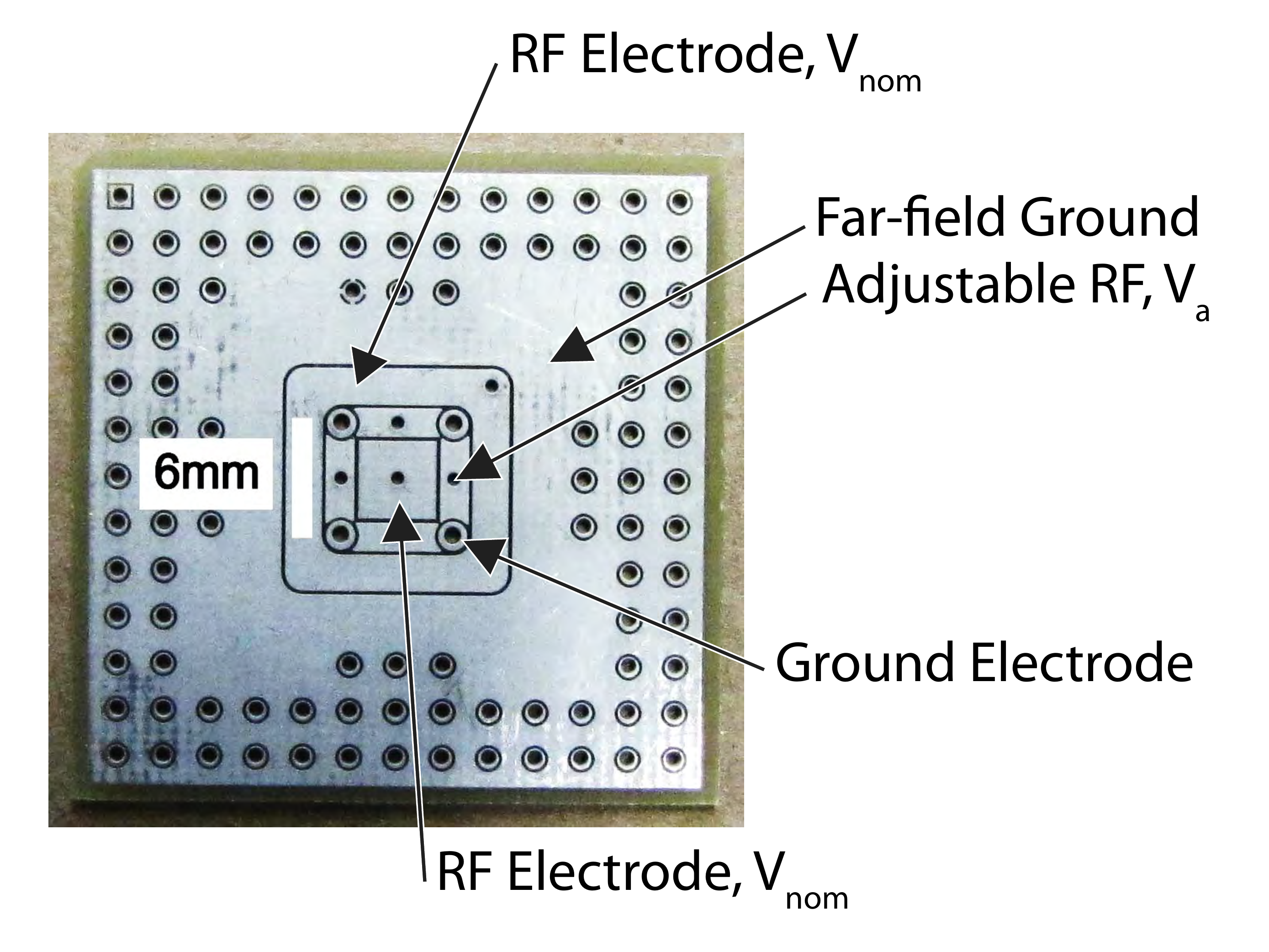}
	\caption{Geometry of the electrodes used for the simulation of trapping \caion ions in a 2$\times$2 array of addressable ion traps as well as for the trapping of charged dust particles.  Above each circular ground electrode is a point Paul trap.  The trap is implemented on a circuit board and has the same physical layout as an industry-standard PGA101 integrated-circuit chip carrier so that it can be plugged into a standard socket.} 	
	\label{fig:dusty}
\end{figure}

The fabricated design is shown in figure~\ref{fig:dusty}.  In between each trapping site is an addressing RF electrode with voltage \vadj, which serves to modify the local trapping potential. This shared electrode allows one to lower the trap frequency of the two neighboring traps.  In the limit of the voltage amplitude going to ground, the two traps merge to become a single linear trap with an axial frequency of almost zero.  Surrounding this 2$\times$2 array is a fixed-voltage-amplitude RF ring electrode at \vnom, and surrounding that is a ground electrode.  To improve \deptheff a ground plane is positioned above the array (not shown in figure~\ref{fig:dusty}).  In a square array the number of addressing electrodes scales as $2N$, where $N$ is the number of point Paul traps.  A hexagonal array would scale as $3N$.  For each addressing electrode, a separate phase-locked RF source is required (detailed in section \ref{sec:trapdrive}).  

\subsection{Simulations of trapping depth and frequency for \caion ions}
\label{sec:dusty-sim}
\begin{figure}
	\centering
	\includegraphics[scale=0.92]{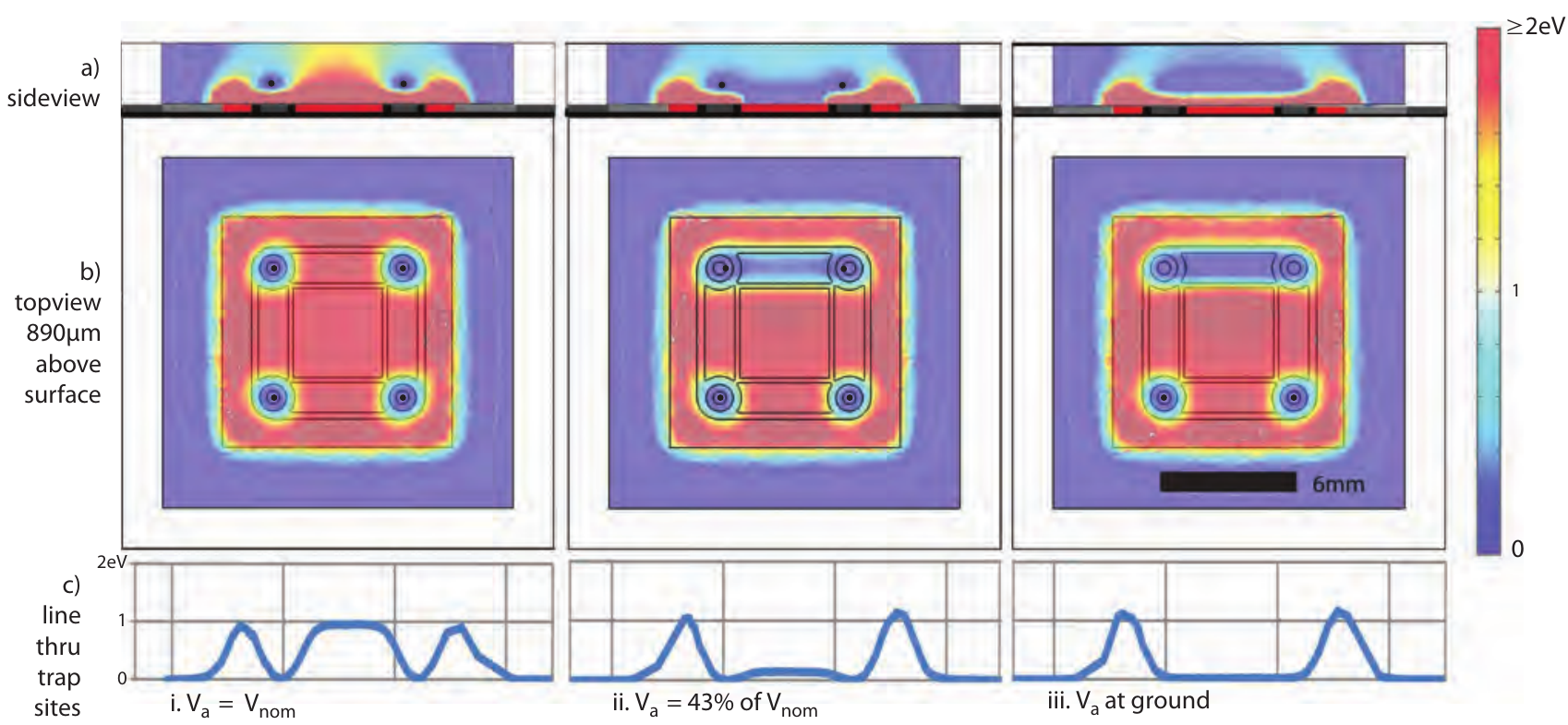} 	
	\caption{Series of simulations for trapping \caion ions in an addressable array of point ion traps with an inter-trap spacing of 6 mm. Top: a) series of simulations showing a slice through the pseudo-potential field in eV through the middle of the two trapping sites which share an addressing electrode. Middle: b) series of simulations showing a slice though the pseudo-potential 890 $\mu$m above the surface of the trap.   Bottom: c) The pseudo-potential along a line connecting the two trapping sites.  From left to right (i-iii), the potential of the addressing electrode \vadj ~is ramped from 100\% of \vnom ~(215 V 10 MHz) to 0 volts.  Initially (i), the two ion traps are fully separated and form two well-isolated point Paul traps.  In between (ii), \vadj ~is 43\% of the nominal trap voltage and the trap frequency has been reduced.  (iii) \vadj ~is at ground and the two addressed ion traps in the array are formed into a linear ion trap.}
	\label{fig:sim-morph}
\end{figure}

Shown in figure~\ref{fig:sim-morph} are the simulation results of the pseudo-potential and trapping frequencies for a 2$\times$2 addressable array of ion traps containing \caion ions with $a$=6~mm inter-trap spacing.  The simulations show that when the addressing RF electrode potential \vadj~is equal to the fixed RF electrodes potential \vnom, the 2$\times$2 array has 4 separate point Paul traps (figure~\ref{fig:sim-morph}i).  As expected, by attenuating \vadj~from 100\% of \vnom~to ground, the two point Paul traps bordering the addressing electrode are morphed into a linear trap.  Because of the asymmetry of the trapping potential as \vadj~is attenuated, the trapping sites move closer to each other (see figure~\ref{fig:sim-morph}ii).  As \vadj~falls, the two trapping sites move toward each other until they are above the boundary of the addressing RF electrode and the ground electrode.  Until the two traps become a linear trap, it is not possible to reduce the distance between the trapping sites further.   The ion interaction could be significantly increased even while maintaining two ions in separate point Paul traps, by making the addressing electrode shorter.  \deptheff would then necessarily be reduced.  The simulations show that a ground plane a distance $a/2$ above the surface of the this planar trap array improves \deptheff from 1.7\% to 6.7\%.  

In the simulations shown here, the addressing electrode length (plus the electrode gap) is about 90\% of  the distance between the home trapping sites, which means that the ions can remain in separate traps, while the distance between them is reduced by $\sim10\%$.  The gate time between point Paul traps scales with the cube of the distance, so that even in this long aspect-ratio trap, the gate time is about 30\% less with addressable electrodes than without.  This is in addition to the increased interaction caused by attenuating the RF voltage, which reduces the secular frequency $\omega$ in both the conventional and addressable arrays of point Paul traps.  The time for this displacement, assuming it is done adiabatically (i.e. without heating the ions), would be approximately 10 times the period of the slowest secular frequency of interest.  Since the secular frequency scales approximately linearly with \vadj, even the slowest traps in table~\ref{tab:gatetimes} could be addressed in microseconds.  A 2D array with 100 $\mu$m inter-trap spacing and 75 $\mu$m long addressing electrodes should be able to perform an entangling gate in 400 $\mu$s, with the adiabatic addressing of the addressing RF electrode taking on the order of 10 $\mu$s.  While the addressing electrodes add significant technical complexity, the increased interaction between neighboring sites and the deeper absolute trapping depth of the whole array means experiments could be significantly easier to perform.

\begin{figure}
\centering
	\includegraphics[scale=0.7]{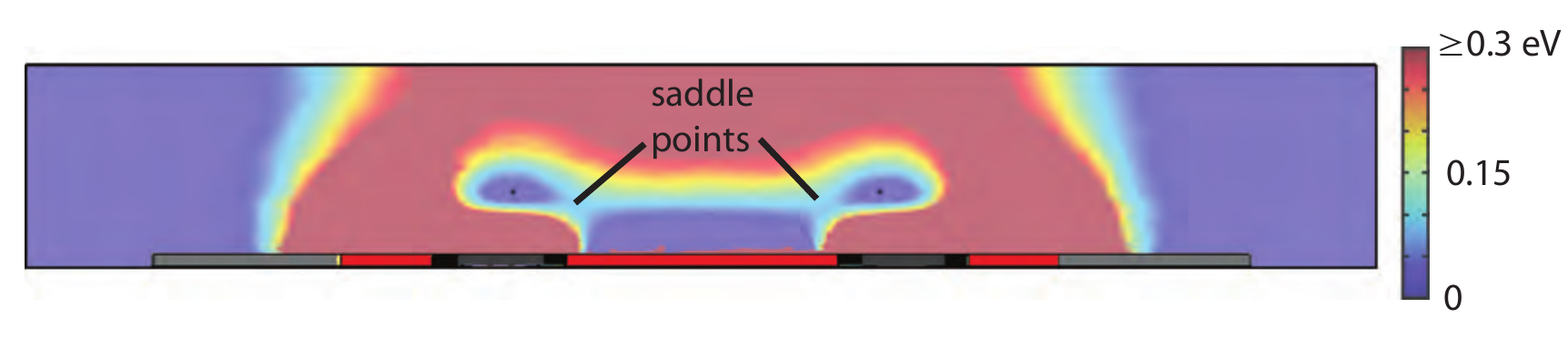}
	\caption{A closer look at the simulation results (from figure~\ref{fig:sim-morph}.a.ii.) when \vadj~is at 43\% of \vnom.  Here the color scale has been set so that the saddle points are easier to see.  A trap appears over the addressing electrode.  Here the trapping locations are indicated by dark points in the spherical trapping potentials.  The ion-ion distance has been also reduced by 10\% compared to the distance with full voltage on the addressing electrode. The potential of the saddle point (here $\sim$0.1~eV) between the new trap over the addressing electrode and the point Paul traps falls with the square of the addressing electrode's voltage, however at all times the ions are trapped within the array by a strong potential.}
	\label{fig:sim-middle}
\end{figure}

When \vadj~is 43\% of \vnom ~(see figure~\ref{fig:sim-middle}), a third trap opens up over the addressing RF electrode.  The height of the saddle points between the two point Paul traps and this third trap falls with the square of \vadj.  However, unlike the trapping potential of a conventional array of point Paul traps with a weak trap depth (see section \ref{sec:gatetimes}), the ions in this array have a strong trapping potential in the overall array.  The addressing electrode allows for the interaction between point Paul traps to be increased while avoiding the problem of a low trapping potential, which leads to ion loss.  As \vadj~is decreased to 0 volts, the third trap rises up and connects the two point Paul traps with a line of RF null.  In this way the two formerly point Paul traps are morphed into a linear Paul trap.  

\section{Results: 2$\times$2 Dust Trap}
\label{sec:dust}
\begin{figure}
	\centering
	\includegraphics[scale=0.7]{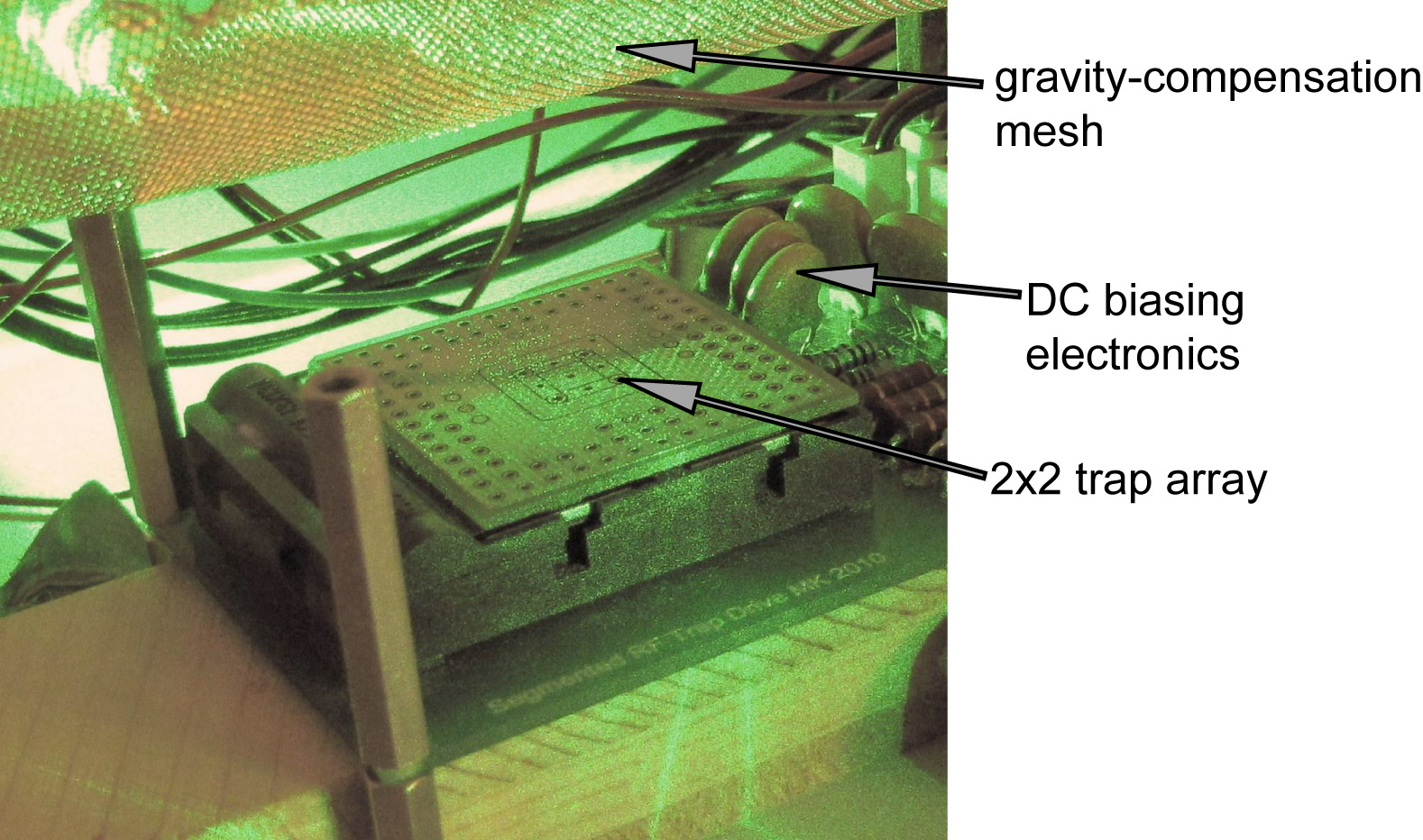}
	\caption{Photograph of the test fixture for trapping charged dust particles.  The RF electrodes are driven with 230 VAC 50 Hz, with the amplitude tuned via auto-transformers.  High-pass filters allow for bias voltages to be applied to all RF electrodes.  The entire assembly is housed in an acrylic box to shield it from air currents.}
	\label{fig:dustysetup}
\end{figure}

A printed circuit board trap (see figure~\ref{fig:dusty}) was used to trap charged dust particles (Lycopodium spores) in air as a proof of principle. The trap was placed in a test fixture inside a plastic acrylic box to shield it from air currents (see figure~\ref{fig:dustysetup}).  The electrodes were driven by auto-transformers, which were AC coupled to the trap to allow for bias DC voltages.  3 cm above the trap was a wire mesh at 150~V, which provided gravity compensation for the charged dust particles.  A green laser pointer was used to illuminate the setup.
\subsection{Trapping and morphing traps with charged dust particles}
\begin{figure}
	\centering
	\includegraphics[scale=0.65]{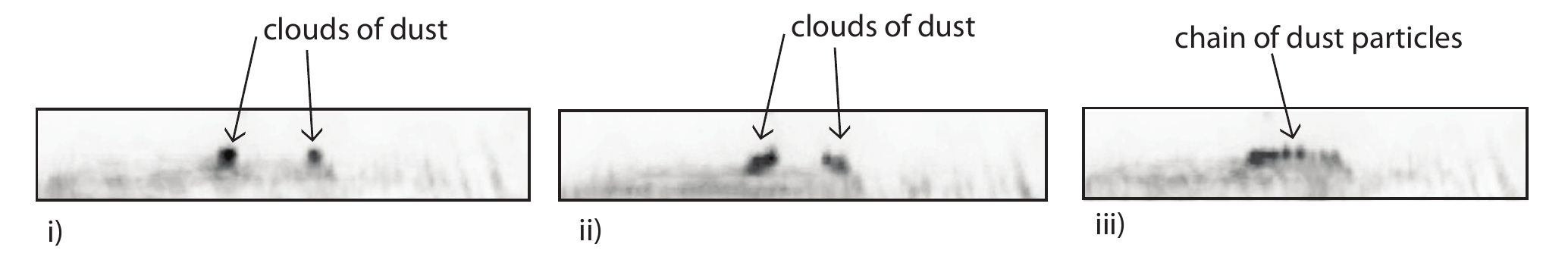}
	\caption{Photographs taken of the charged-particle dust trap.  The voltage on the addressing RF electrode is lowered (pictures i.-iii.) and shows how the two point-like traps are morphed into a single linear trap.  i) Two point Paul traps, each with several dust particles.  ii) The cloud of particles in each separate trap have enlarged and moved toward each other as the two traps are relaxed. iii) linear trap with a chain of dust particles. See on-line supplementary data for a video of this process. }
	\label{fig:dusty-morph}
\end{figure}

By attenuating the voltage on the addressing RF electrode, two point Paul traps could be morphed into a linear trap (see figure~\ref{fig:dusty-morph}).  With full voltage amplitude (\vadj=\vnom=230~VAC) on the addressing RF electrode, each point Paul trap contained a small cloud of charged dust particles.  As the voltage amplitude was lowered, the two traps relaxed towards each other.  When the addressing RF electrode was at ground, the charged dust particles formed into a line.  The parameters used for trapping charged dust particles were different from those needed when trapping ions, because dust particles have a much lower charge to mass ratio.  Instead of a 10~MHz RF drive, just 50~Hz is required.  Despite the low frequency, this is still termed the RF drive in analogy with ion traps.  The secular frequency of the charged particles could be seen and is $\sim$8~Hz.  Because of the small charge to mass ratio of the charged dust particles, they are affected by gravity and require a static field to pull them upwards, so that they are on the RF quadrupole null.
\subsection{Importance of bias voltages on radio frequency electrodes}
The charged dust particles can be steered or shuttled by DC bias voltages on the RF drive electrodes.  When two traps are in a linear trap configuration (figure~\ref{fig:dusty-morph}.iii), a DC bias on an RF electrode can shuttle the charged dust particles from one area of the trap to another.  During the process of morphing the two point Paul traps into a linear trap, a small DC bias voltage ($\sim$2~V) was also used to keep the charged dust particles, which were vibrating with micro-motion, from falling into the third trap which opens up over the addressing RF electrode.  In addition, DC bias voltages can be used to minimize micro-motion caused by the particles being pushed out of the RF quadrupole null by other fields, such as gravity or stray charged particles near the trap.  Because of how useful the DC bias voltages on the electrodes were, it is desirable to also provide this capability in the electrode drive electronics when trapping ions.

\section{Design: 4$\times$4, 16 site ion trap array ``Folsom"}
\label{sec:folsom}
Following a more recent tradition of naming ion-trap geometries after famous prisons, this trap design with addressable RF electrodes is denoted ``Folsom''.  It is designed with the intent to trap up to 16 single \caion ions and allow for the interaction between the inner 2$\times$2 array to be addressed and adjusted.  As explained above, an addressable ion trap array capable of performing quantum gate operations, as described above, between nearest neighbors would need to have a ion-ion spacing $a$ of approximately 100~$\mu$m, which would require electrodes with a feature size of $\sim$25~$\mu$m.  This is possible but technically demanding, especially since it would require many individually controllable electrodes which need to be connected to external sources without obstructing optical access for lasers.  In the first instance, a somewhat larger version has been built to trap ions and test the ideas presented and to develop the technology further.  

\subsection{Folsom geometry and simulations}
\begin{figure}
   \centering
	\includegraphics[scale=1.2]{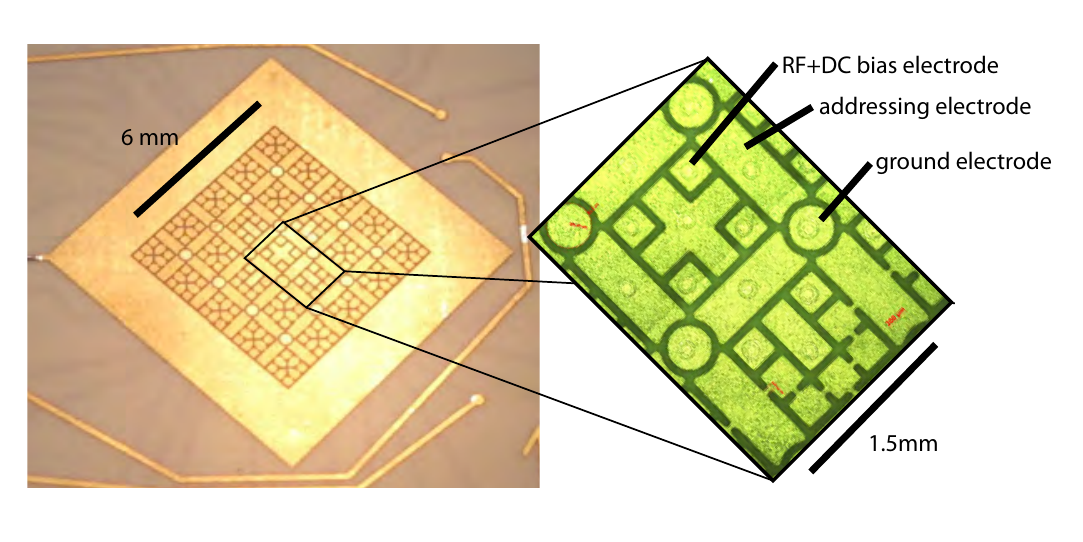}
	\caption{Fabricated trap.  Inside the diamond area, is a 4$\times$4 array of planar point Paul ion traps.   A microscope close-up of the middle 2$\times$2 section of the trap shows details of the electrode structure.  The nominal trapping sites (circular electrodes \diameter400~$\mu$m) are 1.5~mm from each other.}
	\label{fig:folsom}
	\label{fig:folsom-layout}
\end{figure}

To trap \caion ions, a 4$\times$4 square array of addressable ion traps, ``Folsom", was designed.  \caion ions require 396.8 nm laser light for detection and cooling~\cite{CRC:1997}, which can be generated by a commercial frequency-doubled diode laser system.  The array needed to be miniaturized to a 1.5~mm spacing to allow for a 6~mm wide 50~$\mu$m thick sheet of laser light to have the required intensity to fully saturate the ions.  Shown in figure~\ref{fig:folsom-layout} is a photograph of the trap before being put into vacuum.  In order to minimize micro-motion both at the point trapping sites and the linear traps possible between each pair of point Paul traps, as well as to create an axial trapping frequency in the linear traps and to shuttle ions around in 2D, the RF drive was segmented into 26 separate electrodes.  Because of the large number of electrodes (see figure \ref{fig:folsom-layout}) only the inner 2$\times$2 part of the array was wired up to be fully adjustable, while the outer 12 trapping sites provide a periodic boundary condition for the inner array.  Folsom is very similar in geometry to the 2$\times$2 array described in section \ref{sec:dusty-sim},  with the exception of having more trapping sites.  The basic shape of the electrodes is the same, but there are more of them, and the RF electrodes have been further segmented to allow for DC biases to provide micro-motion compensation, to shuttle ions and to impose an axial trapping frequency, when two point Paul traps have been morphed into a linear trap. 

\begin{figure}
   \centering
	\includegraphics[scale=0.6]{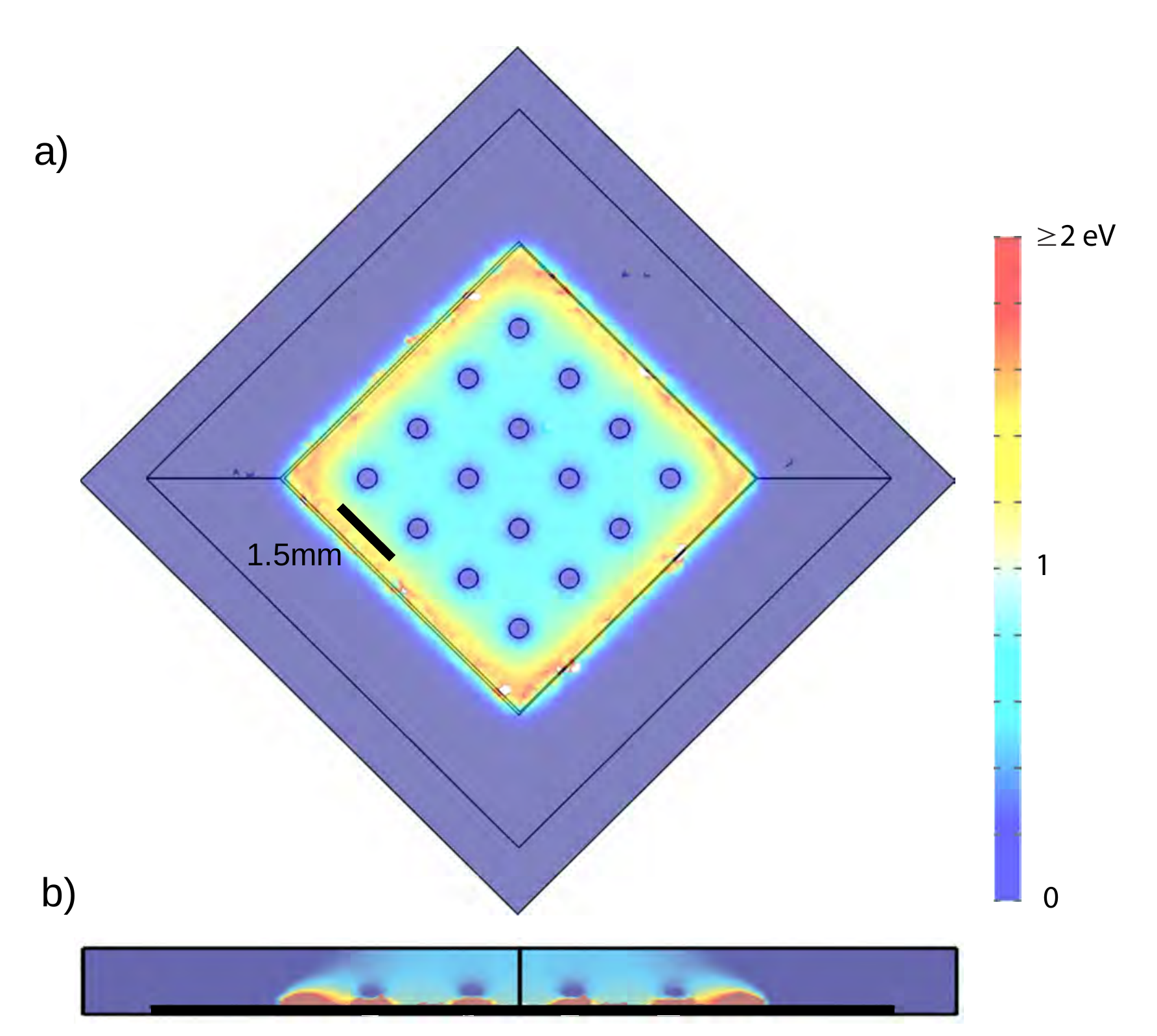}
	\caption{Simulation of Folsom trapping \caion ions showing the pseudo-potential (eV) 375~$\mu$m above the surface of the trap (above) and also from the side diagonally through the trap.  Trapping depth is at least 0.5~eV.  All RF electrodes are fully driven with a voltage amplitude of 125~V and a frequency of 10~MHz.}
	\label{fig:folsom-sim}
\end{figure}

Figure~\ref{fig:folsom-sim} shows a simulation of the pseudo-potential of the 4$\times$4 array.  It is qualitatively similar to the 2$\times$2 array.  With a drive voltage of 125~V at 10~MHz the ions are held within the array with at least 0.5~eV.  A ground plane 1.5~mm above the surface improves \deptheff and should provide shielding from stray charges in the setup.  

\subsection{Material properties and geometry}
The Folsom design was fabricated and is shown in figure \ref{fig:folsom}.  The trap substrate is 35mm$\times$35mm square and 170~$\mu$m thick, made from a vacuum compatible PCB (RO4350b).  The electrodes are 15 ~$\mu$m thick Cu and were patterned using an etching process with a nominal 50~$\mu$m gap between the electrodes.  The board has two layers, where the top side has the trapping electrodes and the bottom has traces which fan-out allowing the electrodes to be connected to the external drive.  Andus GmbH in Berlin produced the circuit board with copper filled vias and a two-layer gold-on-silver plating process (0.13-0.25~$\mu$m Ag, 0.02-0.05~$\mu$m Au). 

\section{Trap drive electronics}
\label{sec:trapdrive}
The simulations described above all assume that the phase of all RF electrodes is the same.  However, if the phases are not the same, the ion location can be significantly altered~\cite{Herskind:2009}.  If the phase is not stable then the ion will experience substantial additional heating.  Because this trap requires many different, individually adjustable, high-voltage, phase-stable RF electrodes, the electronics are detailed below.    The 2$\times$2 fully adjustable array requires at least 5 adjustable high-voltage RF sources, which all maintain the same phase.  Depending on how the DC bias capability is wired up, many more RF sources could also be required. 
\subsection{Tank resonator}
\begin{figure}
\centering
	\includegraphics[scale=1.0]{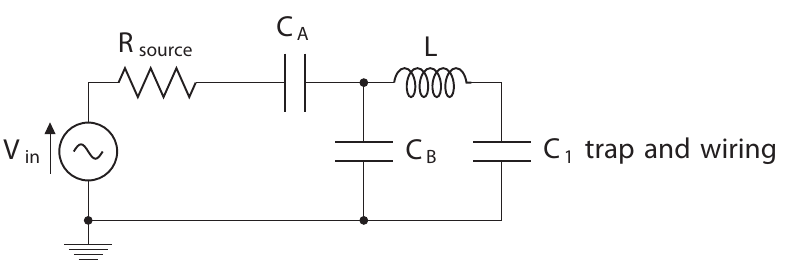}
	\caption{Schematic representation of a tank resonator, which uses an impedance matching network of two capacitors C$_{\mathrm{A}}$ and C$_{\mathrm{B}}$, to match the resonator's impedance to that of the  source impedance R$_{\mathrm{load}}$ of the voltage source V$_{\mathrm{in}}$.}
	\label{fig:tankres}
\end{figure}

In order to create the relatively high voltage (typically $\sim$100 volts) RF signal to drive the ion trap electrodes, a resonant circuit is employed.  Without the resonator, the power to drive an electrode would be $P=V^2/R$, with R=$50$~$\Omega$, or $\sim$200~W.   Many ion traps employ a helical resonator \cite{macalpine:1959}, which can achieve a very high voltage gain (typically $\sim$50) and loaded Q of several hundred.  However they are typically $\sim$10~cm in diameter and $\sim$20~cm long.  Since they must be placed as close to the trap as possible, typically on the vacuum feed-through, and it is necessary to have many resonators in the addressable 2D arrays, a smaller resonator was designed and tested (see figure~\ref{fig:tankres}).

 A common way of viewing the operation of these resonators is as an impedance matching network which matches the impedance of the RF source to the impedance of the electrode~\cite{Abrie:1999}.  Since the electrode and associated wiring is essentially a capacitive load, the RF power necessary to drive the electrodes is only a function of how good of a resonator we can build  (a better resonator requires less power).  This allows for less RF power to be used driving the trap, while it also acts as a band pass filter, which reduces noise on the electrodes.
\subsection{Tank resonator results }
\label{sec:tankres}
An example of this type of tank resonator was built and installed on an existing ion trap experiment and \caion ions were loaded.  The trap was a linear Paul trap with planar trap electrodes and an ion-electrode spacing of 475~$\mu$m. The drive frequency was 10.5~MHz.  A 5~W amplifier was used to drive the resonator.  The elements for the circuit were a 4.7~$\mu$H inductor (API Delevan 4470-09F) with measured unloaded Q=84 at 10.5~MHz, and RF capacitors, C$_{\mathrm{A}}$=220~pF, and C$_{\mathrm{B}}$=820~pF.  The capacitors were chosen to match the source impedance to the load impedance, so as to maximize the voltage gain.  The measured capacitance of the trap, associated wiring and vacuum feed-through was 47~pF.  A capacitive divider was used at the output of the resonator to measure the voltage of the resonator.  The resonator gave a voltage gain of 22.5 and a measured, loaded Q of 51 (resonant frequency divided by FWHM of bandwidth).  This circuit could output 1000~V peak to peak continuously.  To characterize the circuit, we consider the voltage gain G of the resonator,
\begin{equation}
G=\eta \sqrt{\frac{Q}{R}},
\label{eqn:resgain}
\end{equation}
where $\eta$ is equal to 17.4  $\Omega^{1/2}$, Q is equal to 84 and R=50 $\Omega$.  The capacitors were chosen to match the impedance of the resonator to the 50 $\Omega$ source.

Though this resonator loaded ions in a linear Paul trap, an improvement to the matching network of C$_{\mathrm{A}}$ and C$_{\mathrm{B}}$ would be to provide a DC path to either ground or a DC voltage source.  This could be done in various ways, for example by adding an RF choke at the junction of C$_{\mathrm{A}}$ and C$_{\mathrm{B}}$, or by the elimination of C$_{\mathrm{A}}$ (requiring a larger C$_{\mathrm{B}}$).

\subsection{Capacitive coupling between 2 tank resonators}
\begin{figure}
\centering
	\includegraphics[scale=1.0]{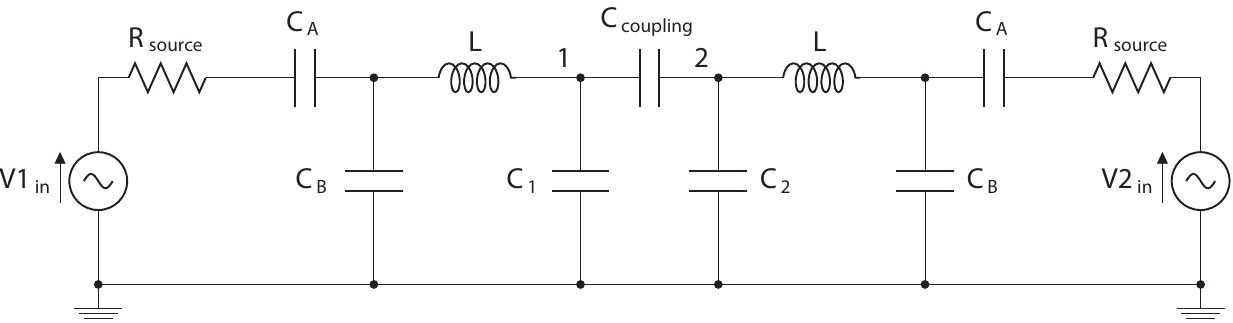}
	\caption{Schematic representation of two coupled tank resonators.  The capacitor between the two trap electrodes C$_{\mathrm{coupling}}$ is the capacitive coupling caused by the physical proximity of the electrodes on the ion trap.  The voltage (amplitude and phase) at node 1 becomes dependent not just on the sinusoidal source V1$_{\mathrm{in}}$, but also the voltage at node 2, leading to instabilities in the drive.}
	\label{fig:coupledres}
\end{figure}

The simple, impedance-matched tank circuit described above works if only one radio frequency drive is required.  If it is necessary to adjust one electrode to a different RF voltage amplitude than that of the others, then it might seem that a scheme involving two resonators and variable sinusoidal sources could be employed (see figure \ref{fig:coupledres}).  

However, the two resonators will be weakly coupled because they drive electrodes which have a capacitive coupling at the trap.  When one electrode voltage is adjusted, it will affect the other resonator and both the phase and amplitude of the RF driving voltages will be unstable.  For example, if both resonators have the same output voltage phase and amplitude at node points 1 and 2 (see figure~\ref{fig:coupledres}) then the coupling capacitor C$_{\mathrm{coupling}}$ does not affect the system.  However, if the second resonator is  set so that node point 2 is at ground, then the capacitance that the first resonator needs to drive will increase by C$_{\mathrm{coupling}}$.  While this is typically very low ($\sim$0.1~pF) it is enough to significantly affect the phase and amplitude of the output of the resonators.  To first order, the change in resonant frequency $\delta f_{0}\propto\delta C/2$, where $\delta C$ is the percentage change in the capacitance of the resonator.  Since the resonators have a loaded Q of $\sim$50 and a load of no more than a few tens of pFs, a change of 0.1~pF would result in phase shifts of $\sim$23 degrees and lead to instabilities in the trap drive.

There are multiple ways to avoid this problem.  One could do away with the resonators, but the required RF power ($\sim$200~W) to drive the ion trap becomes very expensive and technically difficult.  The trap could also be operated with a low-Q resonator~\cite{Karin:2010}, but instabilities in the phase would likely still cause high ion heating and the RF power required ($\sim$10~W) makes scaling the system difficult.  A third way, used here, is to actively lock the resonator.

\subsection{Phase locked tank resonator}
\begin{figure}
\centering
	\includegraphics[scale=1.0]{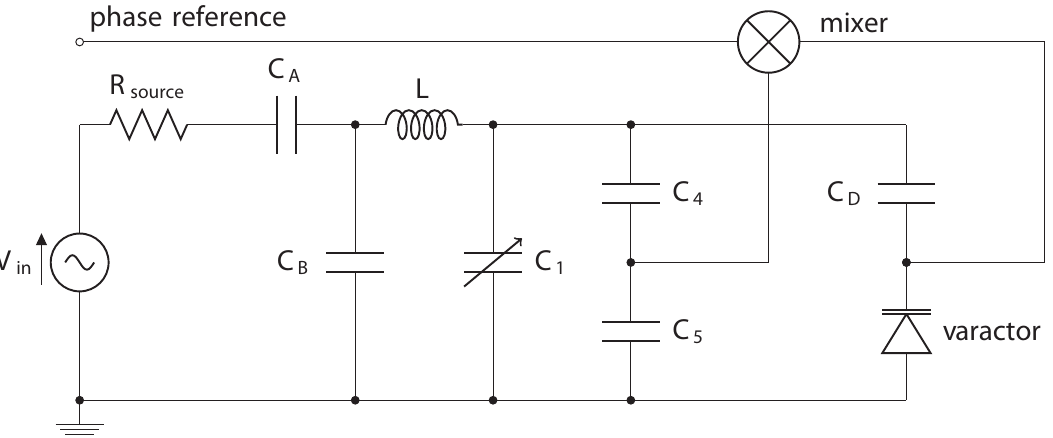}
	\caption{Simplified schematic representation of a phase-locked tank resonator.  With the use of a mixer and an adjustable capacitor (varactor diode), feedback locks the phase and resonance of the resonator even when the trap capacitance (C$_{\mathrm{1}}$) varies due to capacitive coupling to other electrodes.}
	\label{fig:phaselocktankres}
\end{figure}

A varactor diode as a variable capacitor can be used to compensate for the variable capacitance of the trap electrodes, which is caused by their capacitive coupling to other electrodes.  Figure~\ref{fig:phaselocktankres} shows a representation of the circuit.  The output of the resonator is measured with a capacitive divider (C$_{4}$ and C$_{5}$, typically 1~pF and 100~pF respectively) and the phase of this resonator is computed with a mixer (Analog Devices AD8302).  Since a resonator has a phase shift of 90 degrees when it is on resonance, with a linear slope at this frequency, this provides a simple control loop with favorable dynamics to lock the phase of the tank resonator.  Care must be taken to protect the varactor diode from the high voltage output of the resonator.  This can be done by using another capacitive divider (with C$_\mathrm{D}$ typically $\sim$2~pF) to lower the RF voltage the varactor diode sees.  The output of the mixer must be be low-pass filtered, biased, and given the correct gain to properly drive the varactor diode and care must be taken to keep the varactor diode reverse biased (not shown in figure).  Initial tests of this phase-locked resonator allow us to compensate for $\sim$0.2 pF of capacitive coupling at 9.7~MHz with less than 1 degree of phase change utilizing a resonator with the same characteristics as given above in section~\ref{sec:tankres}.  Without the phase-lock, the resonator would have a phase shift of $\sim$30 degrees.

\section{Summary and outlook}
Two-dimensional arrays of addressable ion traps and charged-particle traps are described, where the addressability is tuned via a variable RF electrode in between the trapping sites. Novel methods to drive the variable RF electrodes are presented. The ability to vary the separation between trapping sites on a 2D array is shown using charged dust particles, such that in the limit of maximum interaction, the two former point Paul traps are morphed into a single linear trap.  The above proposal suggests that the same principles can be applied to the 4$\times$4 2D array Folsom and install it in a room temperature vacuum setup to test these ideas with calcium ions.  While this 2-dimensional array currently has a trap spacing too large (1.5~mm) for an entangling gate to be performed between point Paul trapping sites, it should be possible to convert the 2 adjacent point Paul traps into a linear trap and perform ion transport in 2 dimensions.  Also, it is expected that valuable technical knowledge will be obtained in the trapping and experimentation with this 2D array, that will be useful when a smaller array of addressable traps is produced.  An array with a trap spacing of less than 100 $\mu$m with addressable electrodes should be able to produce a fast enough gate that a scalable 2 dimensional array of ions could be used to perform quantum experiments.
\section*{Acknowledgments}
\addcontentsline{toc}{section}{Acknowledgments}
We gratefully acknowledge the support of the European Research Council for the project CRYTERION and the Institute for Quantum Information Ges.m.b.H

\end{document}